\documentstyle[prl,aps,multicol,epsf]{revtex}
\newcommand{\ve}{\varepsilon}
\begin{document}
\draft

\title{Rashba spin splitting in different quantum channels}
\author{Wolfgang H\"ausler}
\address{Fakult\"at f\"ur Physik, Universit\"at
Freiburg, Hermann-Herder-Str.\ 3, 79104 Freiburg, Germany}
\maketitle

\begin{abstract}
Rashba precession of spins moving along a one-subband quantum
channel is calculated. The quantitative influence of unoccupied
higher subbands depends on the shape of the transversal
confinement and can be accounted for perturbatively. Coulomb
interactions are included within the Tomonaga--Luttinger model
with spin-orbit coupling incorporated. Increasing interaction
strength at decreasing carrier density is found to {\sl enhance}
spin precession.
\end{abstract}

\pacs{Keywords: Rashba spin splitting; spin transport; quantum wires;
Tomonaga-Luttinger liquid}
\begin{multicols}{2}
\narrowtext

Complete understanding of the Datta-Das spin transistor
\cite{dattadas} requires to know how different physical
parameters influence the precession of spins which is caused by
the spin-orbit coupling
\begin{equation}\label{hso}
H^{\rm so}=\alpha(\sigma_xp_z-\sigma_zp_x)\;.
\end{equation}
$H^{\rm so}$ breaks spin rotation invariance proportional to
the momentum of spins moving along the `active' $x-z$--plane.
For isotropic single particle energy dispersions $\ve(|k|)$ in
the plane, ignoring Coulomb interactions, this yields the well
known spin split bands $\ve_{\pm}(|k|)=\ve(|k|)\pm\alpha|k|$.
The Rashba parameter $\alpha$ is proportional to the intrinsic
or by means of gates externally applied electric field
perpendicular to the layer \cite{rashba}, here taken as the
$y$--direction. Spin precession occurs then on the length scale
$|k_+-k_-|^{-1}$ of the spin split momenta at the Fermi energy.
It is special to the effective mass approximation
$\ve(|k|)=k^2/2m$, describing many semiconductors, that the
Rashba length $2/|k_+-k_-|=(m\alpha)^{-1}$ does {\em not} depend
on the Fermi energy nor the carrier density.

\begin{figure}
\epsfxsize=0.8\columnwidth
\centerline{\epsffile{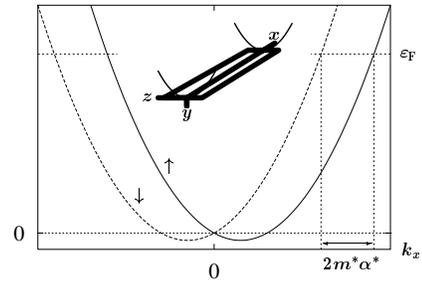}}
\caption[]{
Energy dispersion in the lowest spin split subband of a quantum wire.
On the wire axis the spins $\:s=\uparrow,\downarrow\:$ are polarized
in the plane of the heterostructure parallel to the $z$--direction.
}
\label{figurename}
\end{figure}

Coherent precession of many spins down to the spin selective
drain of the transistor requires to diminish the directional
spread of the electron momenta by introducing quasi
one-dimensional (1D) `wave guides' \cite{dattadas}. To leading
order in $\alpha$ the wave guide simply projects the momenta in
$\ve_{\pm}(|k|)\to\ve_{\pm}(k_x)$ onto the $x$--axes, leaving
the basic features of the 1D case valid, cf.\ Fig.\
\ref{figurename}. In particular will the linear kinetic energy
dispersion in carbon nanotubes or in narrow gap semiconductors
lead to enhanced precession with carrier density.

To higher order $H^{\rm so}$ will mix different transport
channels in each wave guide. Up to O$(\alpha^5)$ this effect
can be accounted for by a renormalizing $\alpha\to\alpha^*$ in
$\ve_{\pm}(k)$ and, within the effective mass approximation, by
$m\to m^*$. In this latter case the Rashba length modifies
according to $(m\alpha)^{-1}\to(m^*\alpha^*)^{-1}$. A
quantitative estimate requires the intra-subband eigenfunctions
\begin{equation}\label{psins}
\psi_{kns}(x,z)={\rm e}^{{\rm i}kx}\phi_n(z)
(\cos(m\alpha z)|s\rangle+{\rm i}\,\sin(m\alpha z)|\!-\!\!s\rangle)\;.
\end{equation}
which are plane waves of momentum $k$ along the wave guide and,
without inter-subband scattering, slightly modified subband
states $\:\phi_n\:$ (subband index $n$) in $z$--direction of the
spin polarization $s=\pm$ on the axes. For a harmonic
confinement (subband energy $\omega_0$), as relevant for example
in gated samples \cite{tarucha}, a perturbative estimate yields
$\:\alpha^*=\alpha(1-\eta)\:$ and $\:m^*=m(1+8\eta^2)\:$ in the
ground subband. Here, the dimensionless parameter
$\eta=(mw\alpha/2)^2$ compares the wave guide width
$w=2/\sqrt{m\omega_0}$ with the Rashba length. For a hard wall
confinement on the other hand (again of width $w$), as possibly
more relevant for wires fabricated by the cleaved edge technique
\cite{yacoby,rother}, the renormalizations become
$\:\alpha^*=\alpha(1-(1/6-1/\pi^2)\eta)\:$ and
$\:m^*=m(1+3(4/3\pi)^6\eta^2)\:$, i.e.\ they are significantly
reduced compared to the soft wall case since $1/6-1/\pi^2\approx
0.065$ and $3(4/3\pi)^6/8\approx 0.002$.

Changing the gate voltage of the transistor when intending to
vary the electrical field and therewith $\alpha$ changes at the
same time the carrier density. Without interactions and when
$\ve(k)=k^2/2m^*$ this would be unimportant. Interactions on
the other hand depend sensitively on the carrier density.
Regarding interaction effects the Tomonaga-Luttinger (TL) model
\cite{tl} provides the most precise low energy description of 1D
metals. Though some of its characteristic power laws can affect
spin properties \cite{egger-epl} we rather focus here on the
question how interactions influence the length over which
coherent spin rotation takes place.

As a second striking property quantum wires exhibit spin--charge
separation which, interestingly and contrary to statements in
the literature \cite{egger-epl,moroz}, is {\em not} spoiled
unless spin-orbit coupling is not exceedingly strong $\eta\sim
1$ if the effective mass description applies. On the other
hand, for non-quadratic dispersion relations, spin charge
separation is in general destroyed, similar as is the case with
Zeeman splitting \cite{frahm-aoki}. An example are carbon
nanotubes where $v_\pm=v_{\rm F}\pm\alpha$ with $\alpha$
originating in this case from the curved surface \cite{cylinder}
instead of the Rashba mechanism.

How to include the Rashba term in the TL-model~? In previous
work \cite{egger-epl,moroz} the Fermi velocities $v_+$ and $v_-$
have been set to different values, which for $\ve(k)=k^2/2m^*$
does not describe $H^{\rm so}$ as can be seen also in Fig.\
\ref{figurename}. Rather {\em both} velocities change slightly
but obey $v_+=v_-$ leading immediately to charge--spin
separation. Thus, in effective mass systems $H^{\rm so}$ acts
solely in the topological spin sector of the corresponding TL
low energy model (of length $L$),
\begin{equation}\label{bosontop}
\frac{\pi}{4L}\left(v_{\mbox{\tiny N}}N_{\sigma}^2+
v_{\mbox{\tiny J}}J_{\sigma}^2\right)-m^*\alpha^*v_{\rm F}
J_{\sigma}\;.
\end{equation}
$N_{\sigma}$ and $J_{\sigma}$ denote the usual currents of
velocities $v_{\mbox{\tiny N/J}}$ where the latter both differ
from $v_{\rm F}$ to account for the Coulomb repulsion
\cite{tl}. In strictly spin isotropic systems $v_{\mbox{\tiny
N}}=v_{\mbox{\tiny J}}$. Since we expect this isotropy being
broken only weakly, both of these velocities should be similar
in magnitude and also similar to the spin velocity $v_\sigma$.
This latter quantity has been determined recently by extensive
quantum Monte--Carlo simulations \cite{creffield}. With
increasing interaction strength, equivalent to a decreasing
carrier density, $v_\sigma/v_{\rm F}$ was found to decrease.
With parameters for existing quantum wires \cite{yacoby}
$v_\sigma/v_{\rm F}$ can drop below 0.5~.

Many quantities of interest can be calculated exactly using
(\ref{bosontop}). In particular it can be shown \cite{hausler}
that spins polarized in $x$--direction along the wire precess
over a length scale
\[
m^*\alpha^*\frac{v_{\sigma{\mbox{\tiny J}}}}{v_{\rm F}}\;.
\]
With $v_{\mbox{\tiny J}}=v_\sigma$ we see that this length
decreases with increasing interaction strength or decreasing
particle density. A similar conclusion has been drawn for
two-dimensional electrons after treating the interactions
perturbatively \cite{raikh}. This trend is opposite to what is
expected for linear single electron dispersions but agrees with
experimental observations \cite{experiments}.

\end{multicols}
\end{document}